\documentclass[preprintnumbers, prd, showpacs, floatfix, preprintnumbers, letterpaper, nofootinbib, amsmath, amssymb, superscriptaddress,preprint]
{revtex4-1}

\usepackage[T1]{fontenc}
\usepackage[latin1]{inputenc}
\usepackage{graphicx}
\usepackage[english]{babel}
\usepackage{amsmath}
\usepackage{amssymb}
\usepackage{amsfonts}
\usepackage{epsfig}
\usepackage{bm}

\usepackage{color}

\begin{document}

\title{Perfect fluid in Lagrangian formulation due to generalized three-form field}

\author{Pitayuth Wongjun}
\affiliation{ The Institute for Fundamental Study,
Naresuan University, Phitsanulok 65000, Thailand}
\affiliation{Thailand Center of Excellence in Physics, Ministry of Education,
Bangkok 10400, Thailand}



\date{\today}

\begin{abstract}
A Lagrangian formulation of perfect fluid due to a noncanonical three-form field is investigated. The thermodynamic quantities such as energy density, pressure and the four velocity are obtained and then analyzed by comparing with the k-essence scalar field. The nonrelativistic matter due to the generalized three-form field with the equation of state parameter being zero is realized while it might not be possible for the k-essence scalar field. We also found that nonadiabatic pressure perturbations can be possibly generated. The fluid dynamics of the perfect fluid due to the three-form field corresponds to the system in which the number of particles is not conserved.We argue that it is interesting to use this three-form field to represent the dark matter for the classical interaction theory between dark matter and dark energy.

\end{abstract}

\maketitle



\section{Introduction}

A theory of cosmological perturbations is one of important issues in cosmology nowadays. It provides us to understand how astronomical structures at large scales are generated and evolve. Also, it can provide us the resulting signatures of the theoretical model to compare with observational data. The theory of cosmological perturbations for a perfect fluid has been developed and studied intensively at the level of equations of motion, for example, a study of the perturbed Einstein field equations together with the equation of conservation of energy momentum tensor \cite{KS,Mukhanov}. Beside the cosmological perturbations at the level of the equations of motion, a study of the cosmological perturbations at the Lagrangian level has been investigated. The advantage point of the study at Lagrangian level is that it is useful to find the perturbed dynamical field as well as derive closed evolution equations. This can be clearly seen by considering the cosmological perturbations in $f(R,G)$ gravity theories where there are two dynamical fields for scalar perturbations \cite{DeFelice:2009ak,DeFelice:2009wp}. For the study in Lagrangian approach, one can straightforwardly identify which fields are dynamical or auxiliary and then immediately obtain the closed evolution equations.

A Lagrangian formulation for a perfect fluid in general relativity has been constructed and developed for a long time \cite{Taub,Schutz,Brown:1992kc}. The Lagrangian of the fluid is simply written as its pressure \cite{Schutz} or energy density \cite{Brown:1992kc}. The advantage point of this formulation is that it naturally provides a consistent way to construct a covariant theory for dark energy and dark matter coupling. The study of dark energy and dark matter coupling has been widely investigated in order to describe a way out from the cosmic coincidence problem \cite{Bettoni:2011fs,Boehmer:2015kta,Boehmer:2015sha,Bettoni:2015wla,Koivisto:2015qua}. Moreover, the observation also provide a hint for the existence of the coupling \cite{Abdalla:2014cla}. However, in order to recover the standard thermodynamics equations, the Lagrangian must involve at least five independent functions. Even though this formulation can provide a consistent way for studying the perfect fluid in cosmology and is well known as a standard approach for the perfect fluid at the Lagrangian level, there might be disadvantage for this approach since the theory involves too many functions.

A simple Lagrangian approach for the perfect fluid has been investigated by using a non-canonical scalar field \cite{Boubekeur:2008kn}, namely k-essence field \cite{ArmendarizPicon:2000dh,ArmendarizPicon:2000ah,Chiba:1999ka}. It was found that the k-essence scalar field can provide a description of the perfect fluid with constant equation of state parameter.  Moreover, it was found that the cosmological perturbations of this kind of the scalar field is equivalent to those in perfect fluid. However, it cannot be properly used to describe a non-relativistic matter with the equation of state parameter being zero since the Lagrangian is not finite. It was also found that the non-adiabatic pressure perturbations cannot be generated \cite{Arroja:2010wy} as well as a vector mode of the perturbations cannot be produced \cite{DeFelice:2009bx}.

Beside the cosmological models due to the scalar field, a three-form field can be successfully used to describe both inflationary models and dark energy models \cite{Germani:2009iq,Koivisto:2009sd,Kobayashi:2009hj,Germani:2009gg,Koivisto:2009fb,Koivisto:2009ew,Ngampitipan:2011se,DeFelice:2012jt,Koivisto:2012xm,Kumar:2014oka,Barros:2015evi}. Even though there is a duality between scalar field and three-form field \cite{Germani:2009iq}, the cosmological models are significantly differed in both background and perturbation levels. At the perturbation level, it is obvious to see that the three-form field can generate intrinsic vector perturbations while it is not possible for the scalar field. Therefore, it might be worthy to find an equivalence between the three-form field with a perfect fluid. In the present work, by mimicking the k-essence scalar field, we consider a generalized version of the three-form field and then find a
possible Lagrangian form to describe the perfect fluid in cosmological background. We found that a simple power-law of the canonical kinetic term can provide the constant equation of state parameter like in the case of k-essence. The advantage point of the three-form field compare with the scalar field is that it can provide a consistent description of the non-relativistic matter field where its equation of state parameter satisfies $w= 0$. The stability issue is also investigated and found that the non-relativistic matter field due to the three-form field is free-from ghost and Laplacian instabilities.

By using the equations of motion of the generalized three-form field, the thermodynamic quantities are identified and found that the perfect fluid due to the three-from field corresponds to fluid in which the number of particles is not conserved. By analyzing the speed of propagation of scalar perturbations and the adiabatic sound speed, we found that the non-adiabatic perturbations can be possibly generated. We argue that it is interesting to use this three-form field to represent the dark matter for the classical interaction theory between dark matter and dark energy.

This paper is organized as follows. In section \ref{sec:model}, we propose a general form of the three-form field and then find the equation of motion as well as the energy momentum tensor. By working in FLRW metric, the energy density and the pressure as well as the equation of state parameter are found. Some specific forms of the Lagrangian satisfying the equations of motion are obtained and found that it can represent the non-relativistic matter. We also investigate the stability issue by using the perturbed action at second order in section \ref{stability}. We found conditions to avoid ghost and Laplacian instabilities. In section \ref{thermo}, we investigate the thermodynamic properties of the model. We begin this section with review of some important idea of the Lagrangian formulation for the standard and k-essence scalar field and then find the thermodynamic properties due to the three-form fluid. Finally, the results are summarized and discussed in section \ref{summary}.

\section{Equations of motion and energy momentum tensor\label{sec:model}}
Cosmological models due to a three-form field have been investigated not only in inflationary models but also dark energy models \cite{Germani:2009iq,Koivisto:2009sd,Kobayashi:2009hj,Germani:2009gg,Koivisto:2009fb,Koivisto:2009ew,Ngampitipan:2011se,DeFelice:2012jt,Koivisto:2012xm,Kumar:2014oka,Barros:2015evi}. Moreover, at the end of inflationary period, a viable model due to the three-form field for the reheating period have  been investigated \cite{DeFelice:2012wy}. A consistent mechanism to generate large scale cosmological magnetic fields by using the three-form field have been studied \cite{Koivisto:2011rm}. Recently, a generalized inflationary model by considering two three-form fields was also investigated \cite{Kumar:2014oka}. All investigations of cosmological models due to three-form are considered only in canonical form. Since the non-canonical form of scalar field have been intensively investigated, it is interesting to investigate the cosmological model with a non-canonical form of the three-form field. In this section, we will consider a non-canonical form of the kinetic term of a three-form  field, $A_{\alpha\beta\gamma}$, as follows
\begin{equation}
S=\int d^{4}x\sqrt{-g}\left[\frac{M_{Pl}^2}{2}R + P(K,y)\right],\label{action-th}
\end{equation}
where the kinetic term and scalar quantity of the three-form field are expressed as
\begin{eqnarray}
K &=&-\frac{1}{48}\,
F_{\alpha\beta\gamma\delta}F^{\alpha\beta\gamma\delta},\\
y&=& \frac{1}{12} A_{\alpha\beta\gamma}A^{\alpha\beta\gamma},\\
F_{\mu\nu\rho\sigma}&=&\nabla_{\mu}A_{\nu\rho\sigma}-\nabla_{\sigma}A_{\mu\nu\rho}+\nabla_{\rho}A_{\sigma\mu\nu}-\nabla_{\nu}A_{\rho\sigma\mu}\,.
\end{eqnarray}
By varying the action with respect to the three-form field, the equations of motion of the three-form field can be written as
\begin{eqnarray}
E_{\alpha\beta\gamma}=\nabla_\mu\left(P_{,K} F^\mu_{\,\,\,\alpha\beta\gamma}\right) + P_{,y}A_{\alpha\beta\gamma} = 0, \label{eom-co}
\end{eqnarray}
where the notation with subscript $P_{,x}$ denotes  $P_{,x} = \partial_x P$. Due to the totally anti-symmetric property of the tensor $F_{\mu\alpha\beta\gamma}$, one found that there exist constraint equations as follows
 \begin{eqnarray}
\nabla_\mu \left( P_{,y}A^{\mu\alpha\beta}\right) = 0. \label{constraint}
\end{eqnarray}
These equations suggest us that the conserved quantity is expressed in terms of three-form field. Note that for the k-essence scalar field, the conserved quantity is expressed in term of one-form or vector quantity. We will discuss on this issue in detail in section \ref{thermo} where we investigate the fluid dynamics. The energy momentum tensor can be obtained by varying the action of the three-form field with respect to the metric as
\begin{eqnarray}
T_{\mu\nu} = \frac{1}{6}P_{,K}F_{\mu\rho\sigma\alpha}F_\nu^{\,\,\,\rho\sigma\alpha}-\frac{1}{2} P_{,y} A_{\mu\rho\sigma}A_{\nu}^{\,\,\,\rho\sigma}+P g_{\mu\nu}. \label{EM-co}
\end{eqnarray}
For consistency of the derived equations, one can check that the conservation of the energy momentum tensor can be obtained up to the equation of motion as follows
\begin{eqnarray}
\nabla_\mu T^{\mu}_{\,\,\nu}=\frac{1}{6} F_{\nu\alpha\beta\gamma}E^{\alpha\beta\gamma} = 0.
\end{eqnarray}

In order to capture the thermodynamics quantities such as the energy density and pressure due to the three-form field like the investigation in scalar field, let us consider a flat Friedmann-Lemaître-Robertson-Walker (FLRW) manifold whose metric element can be written as
\begin{equation}
ds^{2}=-dt^{2}+ \gamma_{ij} dx^i dx^j =-dt^{2}+a(t)^{2}\delta_{ij} dx^i dx^j\,.
\end{equation}
By using this form of the metric and the constraint equation in  Eq.~(\ref{constraint}),
the components of the three-form field, $A_{\alpha\beta\gamma}$, can be written as
\begin{eqnarray}
A_{0ij}=0\,,\qquad A_{ijk}= \epsilon_{ijk} \, X(t)=\sqrt{\gamma}\varepsilon_{ijk} \, X(t) =a^3\varepsilon_{ijk} \, X(t), \label{A-components}
\end{eqnarray}
where $\varepsilon_{ijk}$ is the three-dimensional Levi-Civita symbol with $\varepsilon_{123}=1$. By using this form of the metric, the components of energy momentum tensor can be expressed as
\begin{eqnarray}
T^0_0 &=& P- 2K P_{,K} ,\\
T^i_j &=& (P - 2K P_{,K} - 2 y P_{,y}) \delta^i_j\,.
\end{eqnarray}
By comparing these components of the energy momentum tensor of the three-form to one from the perfect fluid, the energy density and pressure of the three-form can be expressed as
\begin{eqnarray}
\rho &=& 2K P_{,K} -P,\label{rho}\\
p &=& P - 2K P_{,K} - 2 y P_{,y} = -\rho - 2y P_{,y}. \label{pressure}
\end{eqnarray}
Note that we have used $y=X^{2}/2$ and $K = (\dot{X} +3 H X)^2/2$ where $H = \dot{a}/a$ is the Hubble parameter.
From the energy density and the pressure above, the equation of state parameter of the three-form can be written as
\begin{eqnarray}
w =\frac{p}{\rho} =  -1 -\frac{2 y P_{,y}}{\rho}.\label{eos}
\end{eqnarray}
The equation of motion of the three-form field in Eq. (\ref{eom-co}) can be written in flat FLRW background as
\begin{eqnarray}
(2 K P_{,KK} + P_{,K})\dot{K} + 2 K P_{,yK} \dot{y} - 2\sqrt{K\,y} P_{,y}  = 0. \label{eom-FLRW}
\end{eqnarray}
From this point, one can check validity of the derived equations by reducing the general form of the action to the canonical one as setting $P = K -V(y)$. As a result, we found that all equations can be reduced to the canonical one investigated in \cite{Germani:2009iq,Koivisto:2009sd,Kobayashi:2009hj,Germani:2009gg,Koivisto:2009fb,Koivisto:2009ew,Ngampitipan:2011se,DeFelice:2012jt,Koivisto:2012xm,Kumar:2014oka,Barros:2015evi}. Substituting $\rho$ from Eq. (\ref{rho}) into Eq. (\ref{eos}), one obtains
\begin{eqnarray}
2 y P_{,y} + (1+w) 2K P_{,K} = (1+w) P. \label{eq-Pform}
\end{eqnarray}
In order to find the form of $P$, one has to solve this equation. It is useful to solve this equation by considering a simple assumption such as taking the equation of state parameter to be a constant, $w = \text{const}$. By using separation of variable method, the solution can be written as
\begin{eqnarray}
P=P_0 K^{\nu} y^{\mu}, \label{P-con-w}
\end{eqnarray}
where $P_0$ is an integration constant and  $\mu$, $\nu$ are the exponent constants obeying the relation
\begin{eqnarray}
\nu = \frac{1 + w -2\mu}{2(1+w)}, \text{or} \,\,\, w = -1 + \frac{2\mu}{1-2\nu}, \,\,\,\, \nu \neq \frac{1}{2}.
\end{eqnarray}
This form of the solution is very useful since one can interpret the three-form field as a non-relativistic matter or dark matter by setting the equation of state parameter as $w = 0$ while it cannot be properly used for k-essence scalar field case. We will show explicitly why we cannot properly use k-essence scalar field for the non-relativistic matter in section \ref{thermo}. In order to study the covariant coupling form between dark matter and dark energy as suggested from the observation \cite{Abdalla:2014cla}, one can use the three-form as the dark matter with the consistent covariant interaction forms. Moreover, it may be interpreted as dark radiation by setting $w = 1/3$. Note that, in the case of $\nu = 1/2$, it corresponds to the trivial solution since the energy density of the field vanishes. It is important to note that the late-time acceleration of the universe can also be achieved by setting $w = -1$. Even though this may not be distinguished to the cosmological constant at the background level, the cosmological perturbations due to this model of the three-form can be significantly deviated from the model of the cosmological constant.

Since the form of the Lagrangian $P$ is obtained by assuming a constant equation of state parameter, the dark energy model from this three-form field cannot be proposed to solve the coincidence problem.  One may allow the equation of state to be varying in order to overcome this issue. One of interesting solutions is assuming that the equation of state parameter depends on the three-form field $w = w(y)$. In order to solve Eq. (\ref{eq-Pform}) to obtain a suitable form of $P$, one may choose the equation of state parameter such as $w= -1 + \lambda y$, where $\lambda$ is a constant. As a result, the solution can be written as
\begin{eqnarray}
P=P_0 K^{\nu} e^{\frac{(1-2\nu)}{2}\lambda y}. \label{P-nonc-w}
\end{eqnarray}
Naively, it is not difficult to obtain the dynamical dark energy due to the generalized three-form. One can set $\lambda$ be effectively small and find the condition to provide an evolution of $y$  such that it evolves from a large value to a small value. However, since it is not in the canonical form, the theory may be suffered from instabilities. In this work, the stability issue will be investigated in the next section. The investigation of the dark energy model due to the generalized three-form is left in further work.

\section{Stability\label{stability}}
In order to capture the stability conditions of the generalized three-form field, we may consider the perturbations of the field. Since the field minimally couples to the gravity, one has to take into account the metric perturbations. However, for simplicity but useful study, we will investigate the stabilities of the model only in a high-momentum limit. This will capture only some stability conditions. Nevertheless, this includes most of the necessary conditions as found in the canonical three-form field \cite{DeFelice:2012jt}. We leave the full investigation in further work where the cosmological perturbations are taken into account. For this purpose, the metric is held fixed as the Minkowski metric and the three-form field can be written as
\begin{eqnarray}
A_{ijk} &=& \varepsilon_{ijk}(X(t) + \alpha(t,\vec{x})),\\
A_{0ij} &=& \varepsilon_{ijk}(\partial_k\beta(t,\vec{x})  +\beta_k(t,\vec{x}) ),
\end{eqnarray}
where $\alpha$ and $\beta$ are  perturbed scalar fields and $\beta_k$ is a transverse vector obeying the relation $\partial_k \beta^k = 0$. This vector field will be responsible for the intrinsic vector perturbation of the three-form field. For the linear perturbations, the scalar and vector modes are decoupled and then they can be separately investigated. For the scalar modes, by expanding the action up to second order in the field,
the second order action can be written as
\begin{eqnarray}
S^{(2)} &=& \int d^4x \Big( \frac{1}{2}\frac{\dot{Q}^2}{(P_{,K} + 2 K P_{,KK})} - \frac{1}{2} P_{,y} (\partial \beta)^2 +\frac{1}{2}P_{,y} c^2_s \alpha^2\Big), \label{actionO2-scalar}\\
\dot{Q} &=& (P_{,K} + 2 K P_{,KK}) \dot{\alpha} + 2\sqrt{K\,y}P_{,K}P_{,y} \alpha - (P_{,K} + 2 K P_{,KK})\partial^2 \beta,\\
c^2_s &=& 1+\frac{2 y P_{,yy}}{P_{,y}} -\frac{4 K y P_{,Ky}^2}{P_{,y} \left(2 K P_{,KK}+P_{,K}\right)}.\label{sss-scalar}
\end{eqnarray}
One can see that the field $\beta$ is non-dynamical so that one can eliminate it by using  its equation of motion. By applying the Euler-Lagrange equation to the above action, the equation of motion for the field $\beta$ can be written as
\begin{eqnarray}
(P_{,K} + 2 K P_{,KK}) \dot{\alpha} + 2\sqrt{K\,y}P_{,Ky} \alpha - (P_{,K} + 2 K P_{,KK})\partial^2 \beta -P_{,y}\beta= 0,
\end{eqnarray}
From this equation of motion, we can replace the quantity $\dot{Q}$ as $\dot{Q} = P_{,y}\beta$. Note that this equation can be obtained by using the component $(0,i,j)$ of the covariant equation in Eq. (\ref{eom-co}). In order to find the solution for $\beta$, it is convenient to work in Fourier space so that the above equation can be algebraically solved. As a result, by substituting the solution of $\beta$ into the action in Eq. (\ref{actionO2-scalar}), the second order action for the scalar perturbations can be rewritten as
\begin{eqnarray}
S^{(2)} &=& \int dt d^3k\Big( F_1 \dot{\alpha}^2 + F_2 \dot{\alpha}\alpha + F_3 \alpha^2 \Big),
\end{eqnarray}
where
\begin{eqnarray}
F_1 &=& -\frac{P_{,y} \left(2 K P_{,KK}+P_{,K}\right)}{2 \left(k^2\left(2 K P_{,KK}+P_{,K}\right)- P_{,y}\right)},\\
F_2 &=& -\frac{2 \sqrt{K\,y}  P_{,K\,y} P_{,y}}{\left(k^2\left(2 K P_{,KK}+P_{,K}\right)- P_{,y}\right)},\\
F_3 &=& \frac{\left(2 y P_{,yy}+P_y\right) \left(2 k^2 K P_{,KK}+k^2 P_{,K}-P_{,y}\right)-4 k^2 K y P_{,K\,y}^2}{2\left(k^2\left(2 K P_{,KK}+P_{,K}\right)- P_{,y}\right)}.
\end{eqnarray}
As we have discussed above, we will consider the stability conditions at high-momentum limit. Therefore, by taking the limit $k^2\rightarrow \infty$, the second order action becomes
\begin{eqnarray}
S^{(2)} =\int dt d^3k \,k^{-2} (-P_{,y})\Big( \frac{1}{2}  \dot{\alpha}^2 -\frac{1}{2} k^2 c^2_s  \alpha^2 - \frac{1}{2}  m^2_A \alpha^2 \Big). \label{actionO2-k}
\end{eqnarray}
where
\begin{eqnarray}
m^2_A = \frac{d}{dt} \Big( \frac{2 \sqrt{K\,y} P_{,Ky} }{(P_{,K} + 2 K P_{,KK})}\Big)-\frac{4 K\,y\, P_{,Ky}^2 }{(P_{,K} + 2 K P_{,KK})^2} .
\end{eqnarray}
Therefore, the condition to avoid ghost instabilities can be written as
\begin{eqnarray}
P_{,y} < 0.
\end{eqnarray}
This condition can be reduced to the canonical case by taking $P = K -V(y)$, which provides the result as $V_{,y} >0$ consistently with the result in \cite{DeFelice:2012jt}. By finding the equation of motion of $\alpha$ from the action in Eq. (\ref{actionO2-k}), one found that the equation is in the form of massive wave equation of mass $M_A$ propagating with speed $c_s$ defined in Eq. (\ref{sss-scalar}). In order to avoid the Laplacian instability, one requires $c^2_s \geq 0$ leading to the condition
\begin{eqnarray}
 1+\frac{2 y P_{,yy}}{P_{,y}} -\frac{4 K y P_{,Ky}^2}{P_{,y} \left(2 K P_{,KK}+P_{,K}\right)} \geq 0.
\end{eqnarray}
To obtain a clear picture of this condition, one may specify the form of $P$. For the form with constant equation of state parameter, $P = P_0 K^{\nu} y^{\mu}$, the sound speed square can be expressed as $c^2_s = w$. Therefore, the three-form field can be interpreted as the  non-relativistic matter up to a perturbation level since  $c^2_s = 0$ and $w = 0$. Moreover, it is obvious that the non-relativistic matter represented by the generalized three-form field is free from ghost and Laplacian instabilities. Note that the dark energy model with $w< -1/3$ for this form of the Lagrangian is suffered from Laplacian instabilities since the sound speed square is negative.

For another simple form of the Lagrangian with $P= P_0 K^\nu e^{\frac{1-2\nu}{2}\lambda y}$, the sound speed square and the equation of state parameter read $c^2_s = 1+\lambda y$ and $w = -1+\lambda y$.  The no-ghost condition can be expressed as $P_0\lambda (2\nu -1) > 0$. At this point, it is possible to obtain a viable model of dark energy due to the generalized three-form field.

Now we will consider the vector mode of the perturbations by following the same step as in the scalar one. As a result, the second order action for the vector perturbations can be written as
\begin{eqnarray}
S^{(2)} &=& \int d^4x  \Big(  -\frac{1}{2}P_{,y} \beta_i\beta^i\Big).
\end{eqnarray}
From this action, one can see that the vector mode does not propagate at linear level. One has to perform non-linear perturbations in order to find stability behavior of the perturbations. If there are propagating degrees of freedom, it implies that the perturbations are strongly coupled. If the vector modes still do not propagate at non-linear level, it may implies that the symmetry of the background metric does not allow the vector mode to propagate. We leave this investigation for further work. A condition to avoid the instabilities coincides with the condition obtained in scalar mode.

In order to find possibility to obtain non-adiabatic perturbations due to the three-form field, one may find a difference between the speed of propagation of scalar perturbations, $c^2_s$, and the adiabatic sound speed, $c^2_a$. If these two kinds of the sound speed are equal, there are no non-adiabatic perturbations while it provides the possibility to generate non-adiabatic perturbations if they are not equal \cite{Arroja:2010wy}. The speed of propagation of scalar perturbations is found in Eq. (\ref{sss-scalar}). For the adiabatic sound speed, one can derived as follows
\begin{eqnarray}
c^2_a &\equiv& \frac{\dot{p}}{\dot{\rho}} = 1 + 2\frac{(P_{,y} + yP_{,yy})\dot{y} + P_{,Ky} y\dot{K} }{P_{,y}(\dot{y} - 2\sqrt{Ky})},\\
&=& c^2_s + \frac{4\sqrt{Ky}}{P_{,y}(\dot{y} - 2\sqrt{Ky})} \left(P_{y} + yP_{,yy} + \frac{yP_{,Ky}(P_{,y} -2KP_{Ky} )}{(P_{,K} + 2K P_{,KK})}\right).\label{sss-adiabatic}
\end{eqnarray}
Note that the second line of the above equation is obtained by using the equation of motion in Eq. (\ref{eom-FLRW}). From this equation, one can see that the sound speed of scalar perturbations and the adiabatic sound speed  are not generally equal. Therefore, it is possible to generate non-adiabatic perturbations from the generalized three-form field. This is one of advantage points of the generalized three-form field compare with the k-essence scalar field.
It is of interest to find a condition for which $c_s^2$ and $c_a^2$ are the same. From Eq. (\ref{sss-adiabatic}), such a condition can be written as
\begin{eqnarray}
P_{,y} + y P_{,yy} + \frac{yP_{,Ky}(P_{,y} -2KP_{Ky} )}{(P_{,K} + 2K P_{,KK})}=
\partial_K \left(\frac{K P_{,K}}{P_{,y}} \right) = 0.
\end{eqnarray}
Note that  the above equation is obtained by using the definition of the energy density  and pressure expressed in Eq. (\ref{rho}) and Eq. (\ref{pressure}) respectively. By following the calculation in \cite{Arroja:2010wy}, a generic Lagrangian for which $c_s^2$ and $c_a^2$ are the same can be written in the form as
\begin{eqnarray}
P= f(K g(y)),
\end{eqnarray}
where $f$ and $g$ are arbitrary functions. Surprisingly, this formula is exactly the same with the formula obtained in the scalar field case. Note that the Lagrangian forms considered in Eq. (\ref{P-con-w}) and Eq. (\ref{P-nonc-w}) belong to this form.

\section{Fluid dynamics due to three-from field\label{thermo}}
In order to compare the results with the standard description of the fluid dynamics for the perfect fluid, let us briefly review an important concept of the standard version for the fluid dynamics. Since the perfect fluid dynamics due to the non-canonical scalar field or k-essence field has been intensively investigated and interpreted as non-relativistic matter field, for example, in the case of massive gravity theory \cite{Gumrukcuoglu:2015nua, Tannukij:2015wmn}, we will also review some important results of the k-essence scalar field before we discuss further on the three-form field.

\subsection{Standard version and k-essence field}

There are many approaches of the standard version for the perfect fluid Lagrangian. We will use Brown formulation \cite{Brown:1992kc} since it is more useful and has been widely used for recent studies in dark energy and dark matter couplings \cite{Boehmer:2015kta,Boehmer:2015sha,Bettoni:2015wla,Koivisto:2015qua}. The Lagrangian of the perfect fluid can be written in terms of the energy density with Lagrange multipliers as

\begin{eqnarray}
S_m [g_{\mu\nu}, j^\mu, \varphi,s,\alpha_A, \beta_A]= \int d^4x \left( -\sqrt{-g} \,\rho + j^\mu (\varphi_{,\mu} + s \theta_{,\mu} + \beta_A \alpha^A_{,\mu}) \right) ,\label{Lagraigian-pf}
\end{eqnarray}
where $\rho = \rho(n,s)$ is the energy density of the fluid, $n$ is a particle number density, $s$ is an entropy density per particle and $j^\mu$ are components of the particle number flux. The second term which is contracted with $j^\mu$ is the Lagrange multiplier term with the Lagrange multiplier fields $\varphi$, $\theta$ and $\beta_A$ where $\alpha_A$ are the Lagrangian coordinates of the fluid with index $A$ running as $1, 2, 3$. $j^\mu$ can be written in terms of the four-velocity $u^\mu$ of the fluid as
\begin{eqnarray}
j^\mu = \sqrt{-g} \,n\, u^\mu.
\end{eqnarray}
The four-velocity satisfies the relation $u_\mu u^\mu = -1$ where $n = |j|/\sqrt{-g}$ and $|j| = \sqrt{-j^\mu g_{\mu\nu} j^\nu} $. The standard energy momentum tensor of the perfect fluid can be obtained by varying the action with respect to the metric $g_{\mu\nu}$ as
\begin{eqnarray}
T_{\mu\nu} = (\rho+p) u_\mu u_\nu + p \,g_{\mu\nu},\label{EM-pf}
\end{eqnarray}
where $p$ is the pressure of the fluid defined as
\begin{eqnarray}
p \equiv n \frac{\partial \rho}{\partial n} -\rho.
\end{eqnarray}
By varying the action with respect to the Lagrange multiplier fields $\theta$ and $\varphi $, the first law of Thermodynamics and the conservation of the particle number can be obtained respectively \cite{Brown:1992kc} as
\begin{eqnarray}
dp &=& n d\mu - T ds, \\
\partial_\nu j^\nu &=& 0.
\end{eqnarray}
where $T$ is a temperature and $\mu$ is a chemical potential defined as
\begin{eqnarray}
\mu \equiv \frac{\rho + p }{n}.
\end{eqnarray}
From these equations of motion together with the conservation of the energy momentum tensor, $\nabla_\mu T^{\mu\nu} = 0$, all main thermodynamics equations can be obtained. For example, conservation of the entropy density can be obtained by using a projection of the conservation equation of the energy momentum tensor along the fluid flow as follows
\begin{eqnarray}
u_\nu \nabla_\mu T^{\mu\nu} = -\frac{\mu}{\sqrt{-g}} \partial_\nu j^\nu - u^\nu T \partial_\nu s = 0.\label{eom-fluid-Lf}
\end{eqnarray}
From these equations, in the viewpoint of field theory, all main thermodynamics equations can be obtained if one can identify the main thermodynamics quantities in terms of the field such as energy density, pressure, four-velocity and chemical potential which give the form of energy momentum tensor as found in Eq. (\ref{EM-pf}). We will show this procedure for instruction in the case of scalar field.

For the k-essence scalar field, we will follow \cite{Boubekeur:2008kn} in which action of the k-essence field can be written as
\begin{eqnarray}
S_\phi = \int d^4 x \sqrt{-g} P(K_\phi),
\end{eqnarray}
where $K_\phi = -\nabla_\mu\phi\nabla^\mu\phi/2$ is the canonical kinetic term of the scalar field.  The corresponding equations of motion of the scalar field can be expressed as
\begin{eqnarray}
\nabla_\mu \Big(P'\nabla^\mu\phi\Big) = 0,\label{eom-scalar}
\end{eqnarray}
where prime denotes the derivative with respect to $K_\phi$. The energy momentum tensor of the scalar field can be written as
\begin{eqnarray}
T_{\mu\nu} =  P' \nabla_\mu \phi \nabla_\nu\phi + g_{\mu\nu} P.
\end{eqnarray}
By comparing this energy momentum tensor with that in the perfect fluid in Eq. (\ref{EM-pf}), the energy density, pressure and the four-velocity can be identified as
\begin{eqnarray}
\rho_\phi &=&  2 K_\phi P' - P, \\
p_\phi &=& P,\\
u^\mu &=& \frac{\nabla^\mu \phi}{\sqrt{2K_\phi}}
\end{eqnarray}
Therefore, the particle number density can be obtained in order to satisfy the conservation of the particle flux as $n_\phi = \sqrt{2K_\phi} P'$ while the chemical potential reads $\mu_\phi = \sqrt{2K_\phi}$. Therefore, one can check that the equation of motion in Eq. (\ref{eom-scalar}) satisfies the equation of the conservation of the particle flux as follows
\begin{eqnarray}
\sqrt{-g} \nabla_\mu \Big(P'\nabla^\mu\phi\Big) = \partial_\mu\Big(\sqrt{-g} P'\nabla^\mu\phi\Big) = \partial_\mu\Big(\sqrt{-g} n_\phi u^\mu\Big) = \partial_\mu j^\mu_\phi =0.
\end{eqnarray}
As a result, all fluid dynamics equations can be derived by using the results in the standard version. Note that the first law of thermodynamics is adopted for the scalar field while in the case of the standard version, it is obtained from the equation of motion. It is important to note that the conservation of the particle flux does not hold if we generalize the Lagrangian of the scalar field as $P = P(K_\phi, \phi)$ since the equations of motion in Eq. (\ref{eom-scalar}) becomes $ \nabla_\mu \Big(P'\nabla^\mu\phi\Big) = -\partial P/\partial \phi$. This is not so surprisingly since the simple scalar field, such as quintessence field, is also equivalent to the system in which the particle flux is not conserved. This can be explicitly seen by taking $P = K_\phi - V(\phi)$. Note that a particular form the Lagrangian $ P(K_\phi, \phi) = f(K_\phi g(\phi) )$ still provides the conserved particle flux. This is due to a suitable field redefinition to provide the Lagrangian depending only on the kinetic term, $P = P(K_\phi)$ \cite{Arroja:2010wy}.

By taking the equation of state parameter to be constant, the form of the Lagrangian obeys a relation
\begin{eqnarray}
 P (1+w_\phi) = 2w_\phi K_{\phi} P'.
\end{eqnarray}
From this equation, one can  find the exact form of the Lagrangian as
\begin{eqnarray}
P = P_0 K_\phi^{\frac{1+w_\phi}{2w_\phi}},\,\,\,\,\,\, \text{where} \,\,\,\, w_\phi \neq 0.
\end{eqnarray}
It is obviously that one cannot properly use this form of the scalar field to describe the non-relativistic matter since its equation of state parameter is zero, $w = 0$. This is one of drawbacks for the k-essence scalar field. As we have shown before, this does not happen in the case of generalized three-form field.

\subsection{Generalized three-form field}

As we have mentioned, one can find the equivalence between the energy momentum tensor of the three-form and the standard perfect fluid and then identify the fluid quantities such as $\rho, p$ and the four-velocity $u^\mu$ in terms of the three-form field. By using these identifications, one can find the consequent thermodynamics equations of the three-form field as done in the scalar field case. The energy density and the pressure have been identified in Eq. (\ref{rho}) and Eq. (\ref{pressure}) respectively. Now, we will identify the four-velocity of the three-form field by comparing the energy momentum tensor of the perfect fluid in Eq. (\ref{EM-pf}) and the energy momentum tensor of the three-form in Eq. (\ref{EM-co}).  As a result, the relation of the four-velocity and the three-form field can be written as
\begin{eqnarray}
(\rho+p) u_\mu u_\nu = \frac{1}{6}P_{,K}F_{\mu\rho\sigma\alpha}F_\nu^{\,\,\,\rho\sigma\alpha}-\frac{1}{2} P_{,y} A_{\mu\rho\sigma}A_{\nu}^{\,\,\,\rho\sigma}+ (2K P_{,K} + 2y P_{,y}) g_{\mu\nu}. \label{umu1}
\end{eqnarray}
Since $F_{\mu\nu\rho\sigma}$ is a totally symmetric rank-4 tensor in 4-dimensional spacetime, it can be written in terms of a covariant tensor $\epsilon_{\mu\nu\rho\sigma} = \sqrt{-g} \varepsilon_{\mu\nu\rho\sigma}$ where $\varepsilon_{\mu\nu\rho\sigma}$ is the Levi-Civita symbol in four-dimensional spacetime. By using the components of the three-form field in Eq. (\ref{A-components}), the field strength tensor can be written as \begin{eqnarray}
F_{\mu\nu\rho\sigma} = (\dot{X} + 3  H X) \epsilon_{\mu\nu\rho\sigma} = \sqrt{2 K} \epsilon_{\mu\nu\rho\sigma}.
\end{eqnarray}
By using this equation, the first term in the right hand side of Eq. (\ref{umu1}) can be rewritten as
\begin{eqnarray}
\frac{1}{6}P_{,K}F_{\mu\rho\sigma\alpha}F_\nu^{\,\,\,\rho\sigma\alpha} = - 2K P_{,K}  g_{\mu\nu}.
\end{eqnarray}
Substituting this equation into Eq. (\ref{umu1}), one obtains
\begin{eqnarray}
(\rho+p) u_\mu u_\nu &=& -\frac{1}{2} P_{,y} A_{\mu\rho\sigma}A_{\nu}^{\,\,\,\rho\sigma}+  2y P_{,y} g_{\mu\nu}, \nonumber\\
 u_\mu u_\nu &=& \frac{1}{4 y} A_{\mu\rho\sigma}A_{\nu}^{\,\,\,\rho\sigma}- g_{\mu\nu}. \label{umu2}
\end{eqnarray}
One can check that the relation $u_\mu u^\mu = -1 $ valid from this relation. Since the tensor $u_\mu u_\nu$  is constructed from two three-form fields, it plays the role of symmetric rank-2 tensor $S_{\mu\nu}$ instead of outer product of two four-velocity. Therefore, it is not trivial to find the form of the four-velocity of the three-form field. However, one may expect that the four-velocity may relate to the three-form field by the relation of the vector and the three-form in four dimensionality as $u^\mu \propto \epsilon^{\mu \alpha \beta \gamma} A_{\alpha\beta\gamma}$. As a result, the four-velocity of the fluid  can be written in terms of the three-form field as
\begin{eqnarray}
u^\mu =\frac{\epsilon^{\mu \alpha \beta \gamma} A_{\alpha\beta\gamma}}{3!\sqrt{2y}},\label{4-velo-th}
\end{eqnarray}
where the three-form field can be written in terms of the four-velocity as
\begin{eqnarray}
A^{\alpha\beta\gamma} = \sqrt{2y}\epsilon^{\mu \alpha \beta \gamma} u_\mu.
\end{eqnarray}
It is not trivial to find the conserved current density corresponding to three-form field. Actually, there are no conserved quantities obtained from invariance of the action under the shift of the field like the scalar field. However, one may find the conserved quantity from the constraint equation in Eq. (\ref{constraint}) as follows
\begin{eqnarray}
j^{\alpha\beta\gamma} = n^{\mu \alpha \beta \gamma} u_\mu = \sqrt{2y} P_{,y} \epsilon^{\mu \alpha \beta \gamma} u_\mu = P_{,y} A^{\alpha\beta\gamma}.
\end{eqnarray}
From this relation, the conserved quantity is now three-form field instead of vector field and the number density now is four-form field instead of scalar field. This equivalence comes from Hodge duality in four-dimensional spacetime. One may obtained the effective particle number density as
\begin{eqnarray}
n = \sqrt{\frac{n_{\mu \alpha \beta \gamma} n^{\mu \alpha \beta \gamma}}{4!}} = \sqrt{2y} P_{,y}.
\end{eqnarray}
Therefore, the usual particle flux for the three-form field can be written as
\begin{eqnarray}
j^\mu = \sqrt{-g}\, n u^\mu = \sqrt{-g}  P_{,y} \frac{\epsilon^{\mu \alpha \beta \gamma} A_{\alpha\beta\gamma}}{3!}.
\end{eqnarray}
This quantity does not trivially vanish due to the equation of motion in Eq. (\ref{eom-FLRW}). Since $\partial_\mu j^\mu \neq 0$ together with Eq. (\ref{eom-fluid-Lf}), it is inferred that the entropy along the fluid flow is not conserved. The non-conservation of the particle flux for the three-form is due to the fact that the action is not invariant under shift of the field. In the scalar field case, the action is invariant under $\phi \rightarrow \phi + \xi$ where $\xi$ is a constant. For general case of the scalar field with $P_\phi = P_\phi(K_\phi, \phi)$, this symmetry is also broken and then its dynamics will corresponds to the non-conservation of the particle flux like in the three-form case. For the three-form, if we restrict our attention to the case where  $P= P(K)$ which is invariant under shift of the field, the particle number density, $n \propto \rho +p \propto P_{,y}$, will always vanish. Also, the equation of state parameter is always equal to $-1$ which cannot be responsible for the non-relativistic matter.

\subsection{Vector field duality}

In order to complete our analysis, let us consider thermodynamics interpretation in terms of the dual vector field. In four dimensional spacetime, the three-form field is dual to the vector field via the Hodge duality, $A_{\alpha\beta\gamma} = \epsilon_{\alpha\beta\gamma\mu} V^\mu$. By using this duality the kinetic term $K$ and scalar function $y$ of the three-form field can be written in terms of the vector field, $V^\mu$, as
\begin{eqnarray}
K &=& -\frac{1}{48}F^2  = \frac{1}{2} \left(\nabla_\mu V^\mu\right)^2,\label{dual-1}\\
y &=& \frac{1}{12}A_{\alpha\beta\gamma}A^{\alpha\beta\gamma}=-\frac{1}{2}V_\mu V^\mu.\label{dual-2}
\end{eqnarray}
Therefore, the action of the vector field is still in the same form as one in Eq. (\ref{action-th}) where $P=P(K,y)$. However, the dynamical fields are now the vector field and the metric. By varying the action with respect to the vector field, the equation of motion for the vector field can be written as
\begin{eqnarray}
\nabla_\mu\left(P_{,K} \nabla_\rho V^\rho g^{\mu\nu}\right) + P_{,y}V^\nu =0. \label{eom-vector}
\end{eqnarray}
As we have done in the three-form case, the energy momentum tensor for the vector field can be obtained as
\begin{eqnarray}
T_{\mu\nu} = -P_{,y} V_\mu V_\nu + (P-2y P_{,y} -2K P_{,K})g_{\mu\nu}.\label{EM-vector}
\end{eqnarray}
Note that we have used the equation of motion for the vector in Eq. (\ref{eom-vector}) to obtain this form of the energy momentum tensor. One can check that this energy momentum tensor is covariantly conserved up to the equation of motion as we expect. From this form of the energy momentum tensor, it is similar to one for the perfect fluid found in Eq. (\ref{EM-pf}). By comparing $T_{\mu\nu}$ of the vector field in Eq. (\ref{EM-vector}) to one of the perfect fluid in Eq. (\ref{EM-pf}), we can identify the pressure of the vector field as follows
\begin{eqnarray}
p = P-2y P_{,y} -2K P_{,K}.
\end{eqnarray}
This form of the pressure for the vector field coincides with one for the three-form field in Eq. (\ref{pressure}). Now we have to identify the four velocity and the energy density of the vector field. Again, by comparing  $T_{\mu\nu}$ of the vector field to one of the perfect fluid, we found that the four velocity, $u^\mu$ must be proportional to $V^\mu$. Therefore, one can write
\begin{eqnarray}
 u^\mu= \frac{V^\mu}{\sqrt{2y} },
\end{eqnarray}
where the proportional function $\sqrt{2y}$ is obtained by using Eq. (\ref{dual-2}) and relation $u_\mu u^\mu = -1$. Note also that, by using Hodge duality, this four velocity is in the same form with one for the three-form case, Eq. (\ref{4-velo-th}). This suggests that the results obtained in terms of three-form field in previous section are trustable. The energy density can be obtained by evaluating  $\rho=-T^0_0$. As a result, we have
\begin{eqnarray}
\rho = 2K P_{,K} - P.
\end{eqnarray}
By using these thermodynamics quantities, one can obtains the other quantities as done in the same manner in previous section such as $n = -\sqrt{2y}P_{,y}$, $\mu = \sqrt{2y}$ and $j^\mu = \sqrt{-g}\, P_{,y} V^\mu$. Note that we do not need to consider FLRW metric in order to find the thermodynamics quantities in the case of vector field while we do in the case of three-form field.
One can see that all thermodynamics quantities obtained in terms of vector field are the same with the results as found in the three-form field case. These is due to the Hodge duality. As a result, this also implies that the particle number is not conserved as found in three-form field case.

We observe that condition of non-conservation of the entropy density along the fluid flow coincides with the condition of generation of non-adiabatic perturbations even though these conditions come from different approach. The conservation of the entropy density is derived from background equation while non-adiabatic perturbations are properties of the fluid at perturbation level. This argument also hold in both scalar field and three-form field cases. Therefore, this may shed light on the interplay between conserved quantities under shift of the field and non-adiabatic perturbations.

It is important to note that the conservation of the energy momentum tensor of the three-form still valid, $\nabla_\mu T^\mu_\nu =0$. The non-conservation quantities mentioned above are the thermodynamically effective quantities.  As we have mentioned, the useful point of this three-form field is that it can represent the non-relativistic matter field with $w = 0$. Therefore, one may interpret it as dark matter.  This may be useful approach for studies of dark energy and dark matter coupling since one can find the covariant interaction terms at the Lagrangian level and then the resulting closed evolution equations are obtained. This issue is of interest and we leave this detailed investigations for further work.

It is worthwhile to note that the generalized three-form field may be dual to scalar field by introducing some non-minimal couplings to the gravity \cite{Germani:2009iq,Germani:2009gg} or nontrivial term into the Lagrangian. Here, we provide a simple example of the Lagrangian form in which the scalar duality is obtained,
\begin{eqnarray}
\mathcal{L} = P(K,y) + \frac{1}{6} A_{\alpha\beta\gamma} \nabla_\mu F^{\mu\alpha\beta\gamma}.
\end{eqnarray}
The scalar duality may obtained by $F_{\mu\alpha\beta\gamma} = \phi \epsilon_{\mu\alpha\beta\gamma}$ and $A_{\alpha\beta\gamma} = \epsilon_{\alpha\beta\gamma\mu} V^\mu$. Therefore, the kinetic term of the three-form field is proportional to a function of scalar field and then one obtains $P(K,y) = P(\phi, y)$. The additional term is proportional to $V^{\mu} \partial_\mu \phi$. Therefore, one can integrate out the field $V_{\mu}$ which turns out that $V_{\mu} \propto \partial_\mu \phi /P_{,y}$. This provides that $y$ is proportional to the kinetic term of the scalar field and then one obtains the scalar k-essence model as $P(K,y) = P(\phi, y) = P(\phi, X)$, where $X = -(\partial_\mu \phi)^2/2$.

\section{Summary\label{summary}}
A Lagrangian formulation of perfect fluid is a powerful tool to study dynamics of the universe, especially interacting approach between dark energy and dark matter. A general description in this formulation invokes many functions and then it is not easy to handle. A k-essence scalar field can be used to describe the dynamics of the perfect fluid in cosmology. At the background level, even though the k-essence scalar field can be used to describe the perfect fluid with constant  equation of state parameter, it cannot properly used for the non-relativistic matter with $w_\phi=0$. At the perturbation level, the k-essence scalar field cannot provide non-adiabatic perturbations as well as intrinsic vector perturbations.

In the present paper, we propose an alternative way to provide non-adiabatic perturbations and intrinsic vector perturbations by using a generalized three-form field. The investigation is begun with proposing  a general form of the action of the three-form field with a function depending on both the kinetic term and the field, $P=P(K,y)$, similarly to the k-essence scalar field. Equations of motion and energy momentum tensor of the three-form field in covariant form have been calculated. By working in FLRW background, the energy density and the pressure as well as the equation of state parameter are found. For the constant equation of state parameter, an exact form of the Lagrangian reads $P= P_0 K^\nu y^\mu$ where $w = -1 + \frac{2\mu}{1-2\nu}$ and $\nu \neq 1/2$. Therefore, one can set $w=0$ by choosing  proper values of the parameters $\mu$ and $\nu$ and then use the generalized three-form field to represent the non-relativistic matter. For non-constant equation of state parameter, we also point out that it is possible to construct an alternative model of dark energy. The stability analysis of the model is also performed. We found the conditions to avoid ghost and Laplacian instabilities. For the fluid with $w=0$, it is free from ghost and Laplacian instabilities. For some specific model of dark energy, we argue that, to avoid the superluminality, the equation of state parameter must be greater than $-1$. In other words, the viable model of dark energy from the generalized three-form field cannot provide the phantom phase of the universe. Note that the no-ghost condition we found in this paper can be trusted only in the high momentum limit. We leave the full investigation for further work where we investigate the cosmological perturbations and observational constraint. One of important problems found in scalar field quintessence is an quantum mechanical consistency. By considering the quantum fluctuation may alter the classical quintessence potential and then provide an instability of the model \cite{Doran:2002bc}. It is also of interest to study the quantum mechanical consistency for the three-form field model. We leave this investigation for further work.

Thermodynamics properties due to the generalized three-form field are also investigated. It is found that this model corresponds to a system with non-conservation of the particle flux. This leads to a non-conservation of the entropy density along the fluid flow. This is not so surprisingly since many models of dark energy, for example quintessence model, also correspond to the non-conservation of the particle flux. We also found some links between non-conservation of the entropy density along the fluid flow which is a thermodynamically effective quantity at the background level and the generation of non-adiabatic perturbations which is a property of the model at perturbation level. This may shed light on the interplay between conserved quantities under shift of the field and non-adiabatic perturbations. We can argue that this is an useful approach for a study of dark energy and dark matter coupling classically since one can find the covariant interaction terms at the Lagrangian level and then the resulting closed evolution equations are obtained. This issue is of interest and we leave this detailed investigations for further work.

\begin{acknowledgements}
The author is supported by Thailand Research Fund (TRF) through grant TRG5780046. The author would like to thank Khamphee Karwan and Lunchakorn Tannukij for value discussion and comments. The author is deeply grateful to the referees for useful comments on the manuscript. Moreover, the author would like to thank String Theory and Supergravity Group, Department of Physics, Faculty of Science, Chulalongkorn University for hospitality during this work was in progress.
\end{acknowledgements}

\end{document}